\documentclass[twoside]{article}%
\usepackage{amssymb}
\usepackage{amsfonts}
\usepackage{amsmath}
\usepackage{graphicx}%
\providecommand{\U}[1]{\protect\rule{.1in}{.1in}}
\newtheorem{theorem}{Theorem}

\newtheorem{lemma}[theorem]{Lemma}

\newtheorem{remark}[theorem]{Remark}

\newenvironment{proof}[1][Proof]{\noindent\textbf{#1.} }{\ \rule{0.5em}{0.5em}}
\begin{document}

\title{\textbf{Self-Similar Blowup Solutions to the 2-Component Camassa-Holm
Equations}}
\author{M\textsc{anwai Yuen\thanks{E-mail address: nevetsyuen@hotmail.com }}\\\textit{Department of Applied Mathematics, The Hong Kong Polytechnic
University,}\\\textit{Hung Hom, Kowloon, Hong Kong}}
\date{Revised 28-Sept-2010}
\maketitle

\begin{abstract}
In this article, we study the self-similar solutions of the 2-component
Camassa-Holm equations%
\begin{equation}
\left\{
\begin{array}
[c]{c}%
\rho_{t}+u\rho_{x}+\rho u_{x}=0\\
m_{t}+2u_{x}m+um_{x}+\sigma\rho\rho_{x}=0
\end{array}
\right.
\end{equation}
with
\begin{equation}
m=u-\alpha^{2}u_{xx}.
\end{equation}
By the separation method, we can obtain a class of blowup or global solutions
for $\sigma=1$ or $-1$. In particular, for the integrable system with
$\sigma=1$, we have the global solutions:%
\begin{equation}
\left\{
\begin{array}
[c]{c}%
\rho(t,x)=\left\{
\begin{array}
[c]{c}%
\frac{f\left(  \eta\right)  }{a(3t)^{1/3}},\text{ for }\eta^{2}<\frac
{\alpha^{2}}{\xi}\\
0,\text{ for }\eta^{2}\geq\frac{\alpha^{2}}{\xi}%
\end{array}
\right.  ,u(t,x)=\frac{\overset{\cdot}{a}(3t)}{a(3t)}x\\
\overset{\cdot\cdot}{a}(s)-\frac{\xi}{3a(s)^{1/3}}=0,\text{ }a(0)=a_{0}%
>0,\text{ }\overset{\cdot}{a}(0)=a_{1}\\
f(\eta)=\xi\sqrt{-\frac{1}{\xi}\eta^{2}+\left(  \frac{\alpha}{\xi}\right)
^{2}}%
\end{array}
\right.
\end{equation}

where $\eta=\frac{x}{a(s)^{1/3}}$ with $s=3t;$ $\xi>0$ and $\alpha\geq0$ are
arbitrary constants.\newline Our analytical solutions could provide concrete
examples for testing the validation and stabilities of numerical methods for
the systems.

Mathematics Subject Classification (2010): 35B40, 35B44, 35C06, 35Q53

Key Words: 2-Component Camassa-Holm Equations, Shallow Water System,
Analytical Solutions, Blowup, Global, Self-Similar, Separation Method,
Construction of Solutions, Moving Boundary

\end{abstract}

\section{Introduction}

The 2-component Camassa-Holm equations can be expressed in the following form%
\begin{equation}
\left\{
\begin{array}
[c]{c}%
\rho_{t}+u\rho_{x}+\rho u_{x}=0\\
m_{t}+2u_{x}m+um_{x}+\sigma\rho\rho_{x}=0
\end{array}
\right.  \label{2com}%
\end{equation}
with
\begin{equation}
m=u-\alpha^{2}u_{xx}, \label{meq}%
\end{equation}
$x\in R$ and $u=u(x,t)\in R$ is the velocity of fluid and $\rho=\rho
(t,x)\geq0$ is the density of fluid. The constant $\sigma$ is equal to $1$ or
$-1.$ If $\sigma=-1,$ the gravity acceleration points upwards \cite{CLZ},
\cite{C1}, \cite{GY}, \cite{GZ} and \cite{G}. If $\sigma=1,$ there exist some
papers regarding the corresponding models \cite{CI}, \cite{ELY2}, \cite{GZ}
and \cite{GY}. When $\rho\equiv0,$ the system returns to the Camassa-Holm
equation \cite{CH}. The searching of a model equation which can capture
breaking waves and peaked traveling waves is a long-standing open problem
\cite{W}. Here, the Camassa-Holm equation satisfies the above conditions as a
model equation. The searching of peaked traveling waves is motivated by the
wish to discover waves replicating a characteristic for the wave of great
height (waves of largest amplitude), that are exact traveling solutions of the
shallow water equations, whether periodic or solitary \cite{C3}, \cite{T} and
\cite{CE2}. The breaking waves can be understood by solutions which remain
bounded but the slope at some point becomes unbounded in a finite time
\cite{CE0}. Meanwhile, there is an alternative derivation of the Camassa-Holm
equation in \cite{J} and \cite{CL}.

With $\sigma=1,$ the system (\ref{2com}) is integrable \cite{I} and \cite{CH}.
It can be expressed as a compatibility condition of two linear systems (Lax
pair):%
\begin{equation}
\psi_{xx}=(\zeta^{2}\rho^{2}+\zeta m+\frac{1}{4})\psi
\end{equation}
and
\begin{equation}
\psi_{t}=(\frac{1}{2\zeta}-u)\psi_{x}+\frac{1}{2}u_{x}\psi
\end{equation}
with a spectral parameter $\zeta.$\newline It is bi-Hamiltonian, the first
Poisson bracket%
\begin{equation}
\{F_{1},F_{2}\}=-\int\left[  \frac{\delta F_{1}}{\delta m}(m\partial+\partial
m)\frac{\delta F_{2}}{\delta m}+\frac{\delta F_{1}}{\delta m}\rho\partial
\frac{\delta F_{2}}{\delta\rho}+\frac{\delta F_{1}}{\delta\rho}\partial
\rho\frac{\delta F_{2}}{\delta m}\right]  dx
\end{equation}
with the Hamiltonian%
\begin{equation}
H=\frac{1}{2}\int\left(  mu+\rho^{2}\right)  dx
\end{equation}
and the second Poisson bracket%

\begin{equation}
\{F_{1},F_{2}\}=-\int\left[  \frac{\delta F_{1}}{\delta m}(\partial
-\partial^{3})\frac{\delta F_{2}}{\delta m}+\frac{\delta F_{1}}{\delta\rho
}\partial\frac{\delta F_{2}}{\delta\rho}\right]  dx
\end{equation}
with the Hamiltonian%
\begin{equation}
H=\frac{1}{2}\int\left(  u\rho^{2}+u^{3}+uu_{x}^{2}\right)  dx.
\end{equation}
There are two Casimirs:%
\begin{equation}
Mass=\int\rho dx
\end{equation}
and
\begin{equation}
\int mdx.
\end{equation}

In this article, we adopt an alternative approach (method of separation) to
study some self-similar solutions of 2-component Camassa-Holm equations
(\ref{2com}). Indeed, we observe that the isentropic Euler, Euler-Poisson,
Navier-Stokes and Navier-Stokes-Poisson systems are written by:
\begin{equation}
\left\{
\begin{array}
[c]{rl}%
{\normalsize \rho}_{t}{\normalsize +\nabla\cdot(\rho\vec{u})} &
{\normalsize =}{\normalsize 0}\\
{\normalsize \rho\lbrack\vec{u}}_{t}+\left(  {\normalsize \vec{u}\cdot\nabla
}\right)  {\normalsize \vec{u})]+\nabla P} & {\normalsize =-}{\normalsize \rho
\nabla\Phi+vis(\rho,\vec{u})}\\
{\normalsize \Delta\Phi(t,x)} & {\normalsize =\alpha(N)}{\normalsize \rho}%
\end{array}
\right.  \label{Euler-Poisson}%
\end{equation}
where $\alpha(N)$ is a constant related to the unit ball in $R^{N}$:
$\alpha(1)=2$; $\alpha(2)=2\pi$ and for $N\geq3,$%
\begin{equation}
\alpha(N)=N(N-2)Vol(N)=N(N-2)\frac{\pi^{N/2}}{\Gamma(N/2+1)}%
\end{equation}
where $Vol(N)$ is the volume of the unit ball in $R^{N}$ and $\Gamma$ is a
Gamma function. And as usual, $\rho=\rho(t,\vec{x})$ and $\vec{u}=\vec
{u}(t,\vec{x})\in\mathbf{R}^{N}$ are the density and velocity respectively.
$P=P(\rho)=K\rho^{r}$\ is the pressure, the constant $K\geq0$ and $\gamma
\geq1$. And ${\normalsize vis(\rho,\vec{u})}$ is the viscosity
function.\newline We may seek the radial solutions
\begin{equation}
\rho(t,\vec{x})=\rho(t,r)\text{, }\vec{u}=\frac{\vec{x}}{r}V(t,r)=:\frac
{\vec{x}}{r}V\text{,}%
\end{equation}
with $r=\left(  \sum_{i=1}^{N}x_{i}^{2}\right)  ^{1/2}$.\newline By the
standard computation, the Euler equations in radial symmetry can be written in
the following form:%
\begin{equation}
\left\{
\begin{array}
[c]{c}%
\rho_{t}+V\rho_{r}+\rho V_{r}+\dfrac{N-1}{r}\rho V=0\\
\rho\left(  V_{t}+VV_{r}\right)  +K\frac{\partial}{\partial r}\rho^{\gamma}=0.
\end{array}
\right.  \label{eqeq1}%
\end{equation}
For the mass equation in radial symmetry, (\ref{eqeq1})$_{1}$, we well know
the solutions' structure (Lemma 3, \cite{Yuen 2}):%
\begin{equation}
\rho(t,r)=\frac{f(\frac{r}{a(t)})}{a(t)^{N}},\text{ }{\normalsize u(t,r)=}%
\frac{\overset{\cdot}{a}(r)}{a(r)}r{\normalsize .}%
\end{equation}

As the 2-component Camassa-Holm equations, (\ref{2com}), are very similar to
the Euler system (\ref{eqeq1}), in some senses, we can apply the separation
method (\cite{GW}, \cite{M1}, \cite{Li}, \cite{Yuen1}, \cite{Yuen 2}) to the
systems (\ref{2com}). In fact, we can deduce the nonlinear partial
differential equations into much simpler ordinary differential equations. In
this way, we can contribute a new class of self-similar solutions in the
following theorem:

\begin{theorem}
\label{thm:1}We define the function $a(s)$ is the solution of the Emden
equation:
\begin{equation}
\left\{
\begin{array}
[c]{c}%
\overset{\cdot\cdot}{a}(s)-\frac{\xi}{3a(s)^{1/3}}=0\\
a(0)=a_{0}\neq0,\text{ }\overset{\cdot}{a}(0)=a_{1}%
\end{array}
\right.  \text{ } \label{Emden}%
\end{equation}
and
\begin{equation}
f(\eta)=\frac{\xi}{\sigma}\sqrt{-\frac{\sigma}{\xi}\eta^{2}+\left(
\frac{\sigma\alpha}{\xi}\right)  ^{2}}%
\end{equation}
where $\eta=\frac{x}{a(s)^{1/3}}$ with $s=3t;$ $\alpha\geq0,$ $\xi\neq0$,
$a_{0}$ and $a_{1}$ are arbitrary constants.\newline For the 2-compenent
Camassa-Holm equations (\ref{2com}), there exists a family of
solutions:\newline(1) for $\sigma=-1,$\newline(1a) with $\xi<0$ and
$a_{0}>0,$
\begin{equation}
\rho(t,x)=\left\{  \frac{f\left(  \eta\right)  }{a(3t)^{1/3}},\text{
}0\right\}  ,\text{ }u(t,x)=\frac{\overset{\cdot}{a}(3t)}{a(3t)}x.
\label{chch0}%
\end{equation}
The solution (\ref{chch0}) blows up in a finite time $T$.\newline(1b) with
$\xi>0$ and $a_{0}<0,$%
\begin{equation}
\rho(t,x)=\frac{f\left(  \eta\right)  }{a(3t)^{1/3}},\text{ }u(t,x)=\frac
{\overset{\cdot}{a}(3t)}{a(3t)}x. \label{CHCH}%
\end{equation}
The solution (\ref{CHCH}) exists globally;\newline(2) for $\sigma=1,$%
\newline(2a) with $\xi>0$ and $a_{0}>0,$
\begin{equation}
\rho(t,x)=\max\left\{  \frac{f\left(  \eta\right)  }{a(3t)^{1/3}},\text{
}0\right\}  ,\text{ }u(t,x)=\frac{\overset{\cdot}{a}(3t)}{a(3t)}x.
\label{ChCh2}%
\end{equation}
The solution (\ref{ChCh2}) exists globally.\newline(2b) with $\xi<0$ and
$a_{0}<0,$%
\begin{equation}
\rho(t,x)=\frac{f\left(  \eta\right)  }{a(3t)^{1/3}},\text{ }u(t,x)=\frac
{\overset{\cdot}{a}(3t)}{a(3t)}x. \label{chch4}%
\end{equation}
The solution (\ref{chch4}) blows up in a finite time $T$.
\end{theorem}

\section{Separation Method}

First, we can design a nice functional structure for the mass equation:

\begin{lemma}
\label{lem:generalsolutionformasseq}For the $1$-dimensional equation of mass
(\ref{2com})$_{1}$:
\begin{equation}
\rho_{t}+u\rho_{x}+\rho u_{x}=0, \label{massequationspherical}%
\end{equation}
there exist solutions,%
\begin{equation}
\rho(t,x)=\frac{f(\frac{x}{a(3t)^{1/3}})}{a(3t)^{1/3}},\text{ }%
{\normalsize u(t,x)=}\frac{\overset{\cdot}{a}(3t)}{a(3t)}x
\label{generalsolutionformassequation}%
\end{equation}
with the form $f(\eta)\geq0\in C^{1}$ with $\eta=\frac{x}{a(3t)^{1/3}}$, and
$a(3t)>0\in C^{1}.$
\end{lemma}

\begin{proof}
We just plug (\ref{generalsolutionformassequation}) into
(\ref{massequationspherical}) to check:
\begin{align}
&  \rho_{t}+u\rho_{x}+\rho u_{x}\\[0.1in]
&  =\frac{\partial}{\partial t}\left(  \frac{f(\frac{x}{a(3t)^{1/3}}%
)}{a(3t)^{1/3}}\right)  +\frac{\overset{\cdot}{a}(3t)}{a(3t)}x\frac{\partial
}{\partial x}\left(  \frac{f(\frac{x}{a(3t)^{1/3}})}{a(3t)^{1/3}}\right)
+\frac{f(\frac{x}{a(3t)^{1/3}})}{a(3t)^{1/3}}\frac{\partial}{\partial
x}\left(  \frac{\overset{\cdot}{a}(3t)}{a(3t)}x\right) \\[0.1in]
&  =\frac{1}{a(3t)^{1/3+1}}(-\frac{1}{3})\cdot\overset{\cdot}{a}%
(3t)\cdot3\cdot f(\frac{x}{a(3t)^{1/3}})+\frac{1}{a(3t)^{1/3}}\overset{\cdot
}{f}(\frac{x}{a(3t)^{1/3}})\frac{\partial}{\partial t}(\frac{x}{a(3t)^{1/3}%
})\\[0.1in]
&  +\frac{\overset{\cdot}{a}(3t)x}{a(3t)}\frac{\overset{\cdot}{f}(\frac
{x}{a(3t)^{1/3}})}{a(3t)}\frac{\partial}{\partial x}\left(  \frac
{x}{a(3t)^{1/3}}\right)  +\frac{f(\frac{x}{a(3t)^{1/3}})}{a(3t)^{1/3}}%
\frac{\overset{\cdot}{a}(3t)}{a(3t)}\\[0.1in]
&  =-\frac{\overset{\cdot}{a}(3t)f(\frac{x}{a(3t)^{1/3}})}{a(3t)^{1/3+1}%
}+\frac{1}{a(3t)^{1/3}}\overset{\cdot}{f}(\frac{x}{a(3t)^{1/3}})\frac
{x}{a(3t)^{1/3+1}}\frac{1}{3}\dot{a}(3t)3\\[0.1in]
&  +\frac{\overset{\cdot}{a}(3t)x}{a(3t)}\frac{\overset{\cdot}{f}(\frac
{x}{a(3t)^{1/3}})}{a(3t)^{1/3}}\frac{1}{a(3t)^{1/3}}+\frac{f(\frac
{x}{a(3t)^{1/3}})}{a(3t)^{1/3}}\frac{\overset{\cdot}{a}(3t)}{a(3t)}\\[0.1in]
&  =0.
\end{align}
The proof is completed.
\end{proof}

In \cite{DXY} and \cite{Y1}, the qualitative properties of the Emden equation%
\begin{equation}
\left\{
\begin{array}
[c]{c}%
\ddot{a}(s)-\frac{\xi}{a(s)^{N-1}}=0\\
\text{ }a(0)=a_{0}>0,\text{ }\dot{a}(0)=a_{1},
\end{array}
\right.
\end{equation}
where $N\geq2$,\newline were studied. Therefore, the similar local existence
of the Emden equation (\ref{Emden}),
\begin{equation}
\left\{
\begin{array}
[c]{c}%
\ddot{a}(s)-\frac{\xi}{a(s)^{1/3}}=0\\
a(0)=a_{0}\neq0,\text{ }\dot{a}(0)=a_{1},
\end{array}
\right.
\end{equation}
can be proved by the standard fixed point theorem \cite{DXY} and \cite{Y1}. To
additionally show the blowup property of the time function $a(s)$, the
following lemmas are needed.

\begin{lemma}
\label{lemma22}For the Emden equation (\ref{Emden}),%
\begin{equation}
\left\{
\begin{array}
[c]{c}%
\ddot{a}(s)-\frac{\xi}{a(s)^{1/3}}=0\\
a(0)=a_{0}>0,\text{ }\dot{a}(0)=a_{1},
\end{array}
\right.  \text{ } \label{eq124}%
\end{equation}
(1) if $\xi<0$, there exists a finite time $S$, such that $\underset
{s\rightarrow S^{-}}{\lim}a(s)=0.$\newline(2) if $\xi>0$, the solution $a(s)$
globally exists, such that%
\begin{equation}
\underset{s\rightarrow+\infty}{\lim}a(s)=+\infty.
\end{equation}

\end{lemma}

\begin{proof}
(1) For the Emden equation (\ref{eq124}), we can multiply $\dot{a}(s)$ and
then integrate it, as follows:%
\begin{equation}
\frac{\dot{a}(s)^{2}}{2}-\frac{3}{2}\xi a(s)^{2/3}=\theta,\label{rt1}%
\end{equation}
with the constant
\begin{equation}
\theta=\frac{a_{1}^{2}}{2}-\frac{3}{2}\xi a_{0}^{2/3}>0,\label{condition1}%
\end{equation}
for $\xi<0.$\newline The equation (\ref{rt1}) becomes
\begin{equation}
\dot{a}(s)=\pm\sqrt{2\theta+3\xi a(s)^{2/3}}.\label{dadt}%
\end{equation}
We can define the kinetic energy thus:
\begin{equation}
F_{kin}:=\frac{\dot{a}(s)^{2}}{2},
\end{equation}
and the potential energy thus:%
\begin{equation}
F_{pot}=-\frac{3}{2}\xi a(s)^{2/3}.
\end{equation}
The total energy is conserved thus:%

\begin{equation}
\frac{d}{ds}(F_{kin}+F_{pot})=0.
\end{equation}
By the classical energy method for conservative systems (in Section 4.3 of
\cite{LS}), the solution has a trajectory. The time for traveling the orbit
$[0,a_{\sup}]$ can be estimated:%
\begin{equation}
S=\int_{0}^{S}ds\leq2\int_{a_{\inf}}^{a_{\sup}}\frac{da(s)}{\sqrt
{2(\theta+\frac{3\xi a(s)^{2/3}}{2})}}=2\int_{0}^{(\frac{2\theta}{3\xi}%
)^{3/2}}\frac{da(s)}{\sqrt{2(\theta+\frac{3\xi a(s)^{2/3}}{2})}}.\label{hk1}%
\end{equation}
Then we let $G(s):=\sqrt{\frac{-3\xi}{2}}a(s)^{1/3},$ have%
\begin{equation}
dG(s)=\sqrt{\frac{-\xi}{6}}\frac{da(s)}{a(s)^{2/3}},
\end{equation}
and
\begin{equation}
da(s)=\sqrt{\frac{-6}{\xi}}a(s)^{2/3}dG(s)=\frac{2}{-3\xi}\sqrt{\frac{-6}{\xi
}}G(s)^{2}dG(s).
\end{equation}
Then, the equation (\ref{hk1}) becomes:%
\begin{equation}
2\int_{0}^{(\frac{2\theta}{3\xi})^{3/2}}\frac{da(s)}{\sqrt{2(\theta+\frac{3\xi
a(s)^{2/3}}{2})}}=\frac{2\sqrt{2}}{-3\xi}\sqrt{\frac{-6}{\xi}}\int_{0}%
^{\sqrt{\theta}}\frac{G(s)^{2}dG(s)}{\sqrt{(\theta-G(s)^{2})}},
\end{equation}
The integral is solvable exactly by the integration formula 17.11.3 on page 82
of \cite{SLL}:%
\begin{equation}
\int_{0}^{\sqrt{\theta}}\frac{x^{2}dx}{\sqrt{(\theta-x^{2})}}=\left.  \left[
\frac{-x\sqrt{\theta-x^{2}}}{2}+\frac{\theta}{2}\sin^{-1}\left(  \frac
{x}{\sqrt{\theta}}\right)  \right]  \right\vert _{x=0}^{x=\sqrt{\theta}}%
=\frac{\theta\pi}{4}.
\end{equation}
Therefore, we showed that there exists a finite time $S$, such that
$\underset{s->S^{-}}{\lim}a(s)=0$.\newline(2) for the case of $\xi>0$, the
potential function is%
\begin{equation}
F_{pot}=-\frac{3}{2}\xi a(s)^{2/3}.
\end{equation}
As the infimum of $a(s)$ is:%
\begin{equation}
a_{\inf}=(\frac{-2\theta}{3\xi})^{\frac{3}{2}}>0,
\end{equation}
the solution $a(s)$ is uniformly bounded below. The time for achieving the
infimum is finite if $a_{1}<0$:
\begin{equation}
S_{1}=\int_{0}^{S_{1}}ds=\int_{(\frac{-2\theta}{3\xi})^{\frac{3}{2}}}^{a_{0}%
}\frac{da(s)}{\sqrt{2(\theta+\frac{3\xi a(s)^{2/3}}{2})}}<+\infty.
\end{equation}
Therefore, for any constant $a_{1}$, the solution $a(s)$ must increase, after
some finite time. And the time for traveling the interval $[\inf
a(s),+\infty)$ or $[a_{1},+\infty),$ is infinite:%
\begin{equation}
S_{2}=\int_{(\frac{-2\theta}{3\xi})^{\frac{3}{2}}}^{+\infty}\frac{da(s)}%
{\sqrt{2(\theta+\frac{3\xi a(s)^{2/3}}{2})}}=+\infty.
\end{equation}
Therefore, we showed that the solution $a(s)$ globally exists, such that%
\begin{equation}
\underset{s\rightarrow+\infty}{\lim}a(s)=+\infty.
\end{equation}
The proof is completed.
\end{proof}

We may let $b(t)=-a(t),$ in the Emden equation (\ref{Emden}), for $a_{0}<0$:%
\begin{equation}
\left\{
\begin{array}
[c]{c}%
\overset{\cdot\cdot}{a}(s)-\frac{\xi}{3a(s)^{1/3}}=0\\
a(0)=a_{0}<0,\text{ }\overset{\cdot}{a}(0)=a_{1},
\end{array}
\right.  \text{ }%
\end{equation}
have
\begin{equation}
\left\{
\begin{array}
[c]{c}%
\ddot{b}(s)-\frac{\xi}{3b(s)^{1/3}}=0\\
b(0)=-a_{0}>0,\text{ }\dot{b}(0)=-a_{1}.
\end{array}
\right.
\end{equation}
Alternatively, the ordinary differential equation can be rewritten as the
dynamical system:%
\begin{equation}
\left\{
\begin{array}
[c]{c}%
\frac{d}{ds}x(s)=y(s)\\
\frac{d}{ds}y(s)=\frac{\xi}{3x(s)^{1/3}}\\
x(0)=a_{0}\neq0,\text{ }y(0)=a_{1}.
\end{array}
\right.
\end{equation}
It is symmetric about the origin $(0,0)$. Therefore, we can use the above
lemma to drive the corresponding lemma for $a_{0}<0$:

\begin{lemma}
\label{lemma33}For the Emden equation (\ref{Emden}),%
\begin{equation}
\left\{
\begin{array}
[c]{c}%
\ddot{a}(s)-\frac{\xi}{a(s)^{1/3}}=0\\
a(0)=a_{0}<0,\text{ }\dot{a}(0)=a_{1},
\end{array}
\right.
\end{equation}
(1) if $\xi<0$, there exists a finite time $S$, such that $\underset
{s\rightarrow S^{-}}{\lim}a(s)=0.$\newline(2) if $\xi>0$, the solution $a(t)$
globally exists, such that%
\begin{equation}
\underset{s\rightarrow+\infty}{\lim}a(s)=-\infty.
\end{equation}

\end{lemma}

After obtaining the nice structure of solutions
(\ref{generalsolutionformassequation}), we simply use the techniques of
separation of variables (\cite{GW}, \cite{M1}, \cite{Li}, \cite{Yuen1} and
\cite{Yuen 2}), to prove the theorem:

\begin{proof}
[Proof of Theorem \ref{thm:1}]From Lemma \ref{lem:generalsolutionformasseq},
it is clear to see our functions (\ref{chch0}), (\ref{CHCH}), (\ref{ChCh2})
and (\ref{chch4}), fit well into the mass equation, (\ref{2com})$_{1}$, except
for two boundary points.

The second equation of 2-component Camassa-Holm equations (\ref{2com})$_{2}$,
becomes:%
\begin{align}
&  m_{t}+2u_{x}m+um_{x}+\sigma\rho\rho_{x}\\
&  =\left(  u-u_{xx}\right)  _{t}+2u_{x}(u-u_{xx})+u(u-u_{xx})_{x}+\sigma
\rho\rho_{x}.\label{abc}%
\end{align}
As our velocity $u$, in the solutions (\ref{chch0}), (\ref{CHCH}),
(\ref{ChCh2}) and (\ref{chch4}), is a linear flow:
\begin{equation}
u=\frac{\dot{a}(3t)}{a(3t)}x,
\end{equation}
we have
\begin{equation}
u_{xx}=0.
\end{equation}
The equation (\ref{abc}) becomes:
\begin{align}
&  =u_{t}+3u_{x}u+\sigma\rho\rho_{x}\\[0.1in]
&  =\frac{\partial}{\partial t}\left(  \frac{\dot{a}(3t)}{a(3t)}\right)
x+3\left(  \frac{\dot{a}(3t)}{a(3t)}\right)  x\frac{\dot{a}(3t)}{a(3t)}%
+\sigma\frac{f(\frac{x}{a(3t)^{1/3}})}{a(3t)^{1/3}}\left(  \frac{f(\frac
{x}{a(3t)^{1/3}})}{a(3t)^{1/3}}\right)  _{x}\\[0.1in]
&  =\left(  3\frac{\ddot{a}(3t)}{a(3t)}-3\frac{\dot{a}(3t)^{2}}{a(3t)^{2}%
}\right)  x+3\left(  \frac{\dot{a}(3t)}{a(3t)}\right)  \frac{\dot{a}%
(3t)}{a(3t)}x+\sigma\frac{f(\frac{x}{a(3t)^{1/3}})}{a(3t)^{1/3}}\frac{\dot
{f}(\frac{x}{a(3t)^{1/3}})}{a(3t)^{1/3}}\frac{1}{a(3t)^{1/3}}\\[0.1in]
&  =3\frac{\ddot{a}(3t)}{a(3t)}x+\sigma\frac{f(\frac{x}{a(3t)^{1/3}})\dot
{f}(\frac{x}{a(3t)^{1/3}})}{a(3t)}\\[0.1in]
&  =\frac{\sigma}{a(3t)}\left(  \frac{\xi}{\sigma}\eta+f(\eta)\dot{f}%
(\eta)\right)  ,
\end{align}
with the Emden equation:%
\begin{equation}
\left\{
\begin{array}
[c]{c}%
\ddot{a}(s)-\frac{\xi}{3a(s)^{1/3}}=0\\
a(0)=a_{0}\neq0,\text{ }\dot{a}(0)=a_{1},
\end{array}
\right.
\end{equation}
by defining the variables $s:=3t$ and $\eta:=x/a(s)^{1/3}.$\newline Now, we
can separate the partial differential equations into two ordinary differential
equations. Then, we only need to solve for $\frac{\xi}{\sigma}<0,$%
\begin{equation}
\left\{
\begin{array}
[c]{c}%
\frac{\xi}{\sigma}\eta+f(\eta)\dot{f}(\eta)=0\\
f(0)=-\alpha\leq0;
\end{array}
\right.  \label{ode1}%
\end{equation}
or for $\frac{\xi}{\sigma}>0,$%
\begin{equation}
\left\{
\begin{array}
[c]{c}%
\frac{\xi}{\sigma}\eta+f(\eta)\dot{f}(\eta)=0\\
f(0)=\alpha\geq0;
\end{array}
\right.  \label{ode2}%
\end{equation}
The ordinary differential equations (\ref{ode1}) or (\ref{ode2}) can be solved
exactly as
\begin{equation}
f(\eta)=\frac{\xi}{\sigma}\sqrt{\frac{-\sigma}{\xi}\eta^{2}+\left(
\frac{\sigma\alpha}{\xi}\right)  ^{2}}.
\end{equation}
In fact, we have the self-similar solutions in details:\newline(1) for
$\sigma=-1,$\newline(1a) with $\xi<0$ and $a_{0}>0$:
\begin{equation}
\left\{
\begin{array}
[c]{c}%
\rho(t,x)=\left\{
\begin{array}
[c]{c}%
\frac{f\left(  \eta\right)  }{a(3t)^{1/3}},\text{ for }\eta^{2}<-\frac
{\alpha^{2}}{\xi}\\
0,\text{ for }\eta^{2}\geq-\frac{\alpha^{2}}{\xi}%
\end{array}
\right.  ,\text{ }u(t,x)=\frac{\overset{\cdot}{a}(3t)}{a(3t)}x\\
\overset{\cdot\cdot}{a}(s)-\frac{\xi}{3a(s)^{1/3}}=0,\text{ }a(0)=a_{0}%
>0,\text{ }\overset{\cdot}{a}(0)=a_{1}\\
f(\eta)=-\xi\sqrt{\frac{1}{\xi}\eta^{2}+\left(  \frac{1}{\xi}\alpha\right)
^{2}}.
\end{array}
\right.  \label{eqeqeq1}%
\end{equation}
(1b) with $\xi>0$ and $a_{0}<0$:%
\begin{equation}
\left\{
\begin{array}
[c]{c}%
\rho(t,x)=\frac{f\left(  \eta\right)  }{a(3t)^{1/3}},\text{ }u(t,x)=\frac
{\overset{\cdot}{a}(3t)}{a(3t)}x\\
\overset{\cdot\cdot}{a}(s)-\frac{\xi}{3a(s)^{1/3}}=0,\text{ }a(0)=a_{0}%
<0,\text{ }\overset{\cdot}{a}(0)=a_{1}\\
f(\eta)=-\xi\sqrt{\frac{1}{\xi}\eta^{2}+\left(  \frac{1}{\xi}\alpha\right)
^{2}}.
\end{array}
\right.  \label{new1}%
\end{equation}
(2) for $\sigma=1,$\newline(2a) with $\xi>0$ and $a_{0}>0$:
\begin{equation}
\left\{
\begin{array}
[c]{c}%
\rho(t,x)=\left\{
\begin{array}
[c]{c}%
\frac{f\left(  \eta\right)  }{a(3t)^{1/3}},\text{ for }\eta^{2}<\frac
{\alpha^{2}}{\xi}\\
0,\text{ for }\eta^{2}\geq\frac{\alpha^{2}}{\xi}%
\end{array}
\right.  ,\text{ }u(t,x)=\frac{\overset{\cdot}{a}(3t)}{a(3t)}x\\
\overset{\cdot\cdot}{a}(s)-\frac{\xi}{3a(s)^{1/3}}=0,\text{ }a(0)=a_{0}%
>0,\text{ }\overset{\cdot}{a}(0)=a_{1}\\
f(\eta)=\xi\sqrt{-\frac{1}{\xi}\eta^{2}+\left(  \frac{1}{\xi}\alpha\right)
^{2}}.
\end{array}
\right.  \label{abc1}%
\end{equation}
(2b) with $\xi<0$ and $a_{0}<0$:%
\begin{equation}
\left\{
\begin{array}
[c]{c}%
\rho(t,x)=\frac{f\left(  \eta\right)  }{a(3t)^{1/3}},\text{ }u(t,x)=\frac
{\overset{\cdot}{a}(3t)}{a(3t)}x\\
\overset{\cdot\cdot}{a}(s)-\frac{\xi}{3a(s)^{1/3}}=0,\text{ }a(0)=a_{0}%
<0,\text{ }\overset{\cdot}{a}(0)=a_{1}\\
f(\eta)=\xi\sqrt{-\frac{1}{\xi}\eta^{2}+\left(  \frac{1}{\xi}\alpha\right)
^{2}}.
\end{array}
\right.  \label{abc2}%
\end{equation}
With the assistance of Lemmas \ref{lemma22} and \ref{lemma33}, the blowup or
global existence of the solutions can be determined under the prescribed
conditions in the theorem.\newline The proof is completed.
\end{proof}

\begin{remark}
For $\xi<0$, the blowup solutions (\ref{chch0}) and (\ref{chch4}) collapse at
the origin%
\begin{equation}
\underset{t\rightarrow T^{-}}{\lim}\rho(t,0)=+\infty,
\end{equation}
with a finite time $T;$\newline For $\xi>0$, the global behavior of the
solutions (\ref{CHCH}) and (\ref{ChCh2}) at the origin is:
\begin{equation}
\underset{t\rightarrow+\infty}{\lim}\rho(t,0)=0.
\end{equation}

\end{remark}

\begin{remark}
The solutions (\ref{chch0}) and (\ref{ChCh2}) are only $C^{0}$ functions, as
the function $f(\eta)$ is discontinuous at the two boundary points, for
$\alpha>0$:%
\begin{equation}
\underset{\eta^{2}\rightarrow\left\vert \frac{\alpha}{\xi}\right\vert }{\lim
}\dot{f}(\eta)\neq0.
\end{equation}

\end{remark}

\begin{remark}
Our analytical solutions could provide concrete examples for testing the
validation and stabilities of numerical methods for the systems. Additionally,
our special solutions can shed some light on understanding of evolutionary
pattern of the systems.
\end{remark}

\begin{remark}
For the integrable system with $\sigma=1$, we may calculate the mass
of\newline(1) the blowup solution (\ref{chch4}) (or (\ref{abc2})), $\xi<0$ and
$a_{0}<0$:%
\begin{equation}
Mass=\int_{-\infty}^{+\infty}\rho(0,x)dx=\frac{\xi}{a_{0}^{1/3}}\int_{-\infty
}^{+\infty}\sqrt{-\frac{1}{\xi}\left(  \frac{x}{a_{0}^{1/3}}\right)
^{2}+\left(  \frac{1}{\xi}\alpha\right)  ^{2}}dx=+\infty;
\end{equation}
(2) the global solution (\ref{ChCh2}) (or (\ref{abc1})), $\xi>0$ and $a_{0}%
>0$:%
\begin{align}
Mass  &  =\int_{-\infty}^{+\infty}\rho(0,x)dx\\[0.1in]
&  =2\int_{0}^{a_{0}^{1/3}\alpha\sqrt{\frac{1}{\xi}}}\frac{f\left(  \frac
{x}{a_{0}^{1/3}}\right)  }{a_{0}^{1/3}}dx\\
&  =\frac{2\xi}{a_{0}^{1/3}}\int_{0}^{a_{o}^{\frac{1}{3}}\alpha\sqrt{\frac
{1}{\xi}}}\sqrt{-\frac{1}{\xi}\left(  \frac{x}{a_{0}^{1/3}}\right)
^{2}+\left(  \frac{\alpha}{\xi}\right)  ^{2}}dx\\[0.1in]
&  =2\xi\int_{0}^{\alpha\sqrt{\frac{1}{\xi}}}\sqrt{-\frac{1}{\xi}s^{2}+\left(
\frac{\alpha}{\xi}\right)  ^{2}}ds\\[0.1in]
&  =\frac{2\xi}{\sqrt{\xi}}\int_{0}^{\alpha\sqrt{\frac{1}{\xi}}}\sqrt
{\frac{\alpha^{2}}{\xi}-s^{2}}ds\\[0.1in]
&  =\frac{\alpha^{2}\pi}{2\sqrt{\xi}}.
\end{align}

\end{remark}

\section{Blowup Rates}

We are interested in how fast the blowup solutions tend to the infinite as the
time tends to the critical time $T$. The blowup rate of the constructed
solutions (\ref{chch0}) and (\ref{chch4}) at the origin, is estimated in the
following theorem:

\begin{theorem}
The blowup rate of the solutions (\ref{chch0}) and (\ref{chch4}) for
$\alpha>0$, is%
\begin{equation}
\underset{s\rightarrow S^{-}}{\lim}\rho(s,0)(S-s)^{1/3}\geq O(1).
\end{equation}

\end{theorem}

\begin{proof}
We only need to study the blowup rate of the Emden equation (\ref{Emden}),%
\begin{equation}
\left\{
\begin{array}
[c]{c}%
\ddot{a}(s)-\frac{\xi}{a(s)^{1/3}}=0\\
a(0)=a_{0}>0,\text{ }\dot{a}(0)=a_{1}%
\end{array}
\right.  \text{ }%
\end{equation}
with $\xi<0.$\newline For the total energy $\theta>0,$ with the assistance of
the equation (\ref{dadt}),%
\begin{equation}
S=s+\int_{s}^{S}d\eta=s+\int_{a(s)}^{0}\frac{d\eta}{da}da=s-\int_{a(s)}%
^{0}\frac{1}{\sqrt{2\theta+3\xi a(\eta)^{2/3}}}da\geq s+\int_{0}^{a(s)}%
\frac{da(\eta)}{\sqrt{2\theta}}.
\end{equation}
That is
\begin{equation}
(S-s)^{1/3}\geq O(1)a(s)^{1/3}.
\end{equation}
Therefore, we may estimate the blowup rates at the origin of the density
function $\rho(s,x)$:%
\begin{equation}
\underset{s\rightarrow S^{-}}{\lim}\rho(s,0)(S-s)^{1/3}=\underset{s\rightarrow
S^{-}}{\lim}\frac{\alpha}{a(s)^{1/3}}(S-s)^{1/3}\geq O(1).
\end{equation}
The proof is completed.
\end{proof}

We notice that our blowup rate in the above theorem is different form the
results of the Constantin and Escher's papers \cite{CE} and \cite{C2}.

\section{Acknowledgement}

The author thanks the reviewers for their helpful comments to improve the
readability of this article.


\begin{thebibliography}{99}                                                                                               %


\bibitem {CH}R. Camassa and D. D. Holm, \textit{Intergrable Shallow Water
Equation with Peaked Solitons}, Phys. Rev. Lett. \textbf{71} (1993), 1661--1664.

\bibitem {CLZ}M. Chen, S.-Q. Liu, Y. Zhang, \textit{A 2-component
Generalization of the Camassa--Holm Equation and Its Solutions}, Lett. Math.
Phys. \textbf{75} (2006), 1--15.

\bibitem {C1}A. Constantin, \textit{On the Blow-up Solutions of a Periodic
Shallow Water Equation}, J. Nonlinear Sci. \textbf{10} (2000), 391--399.

\bibitem {C2}A. Constantin, \textit{Existence of Permanent and Breaking Waves
for a Shallow Water Equation: a Geometric Approach,} Ann. Inst. Fourier
(Grenoble) \textbf{50} (2000), 321--362.

\bibitem {C3}A. Constantin, \textit{The Trajectories of Particles in Stokes
Waves}, Invent. Math. \textbf{166} (2006), 523--535.

\bibitem {CE0}A. Constantin and J. Escher, \textit{Wave Breaking for Nonlinear
Nonlocal Shallow Water Equations}, Acta Math. \textbf{181} (1998), 229--243.

\bibitem {CE}A. Constantin and J. Escher, \textit{On the Blow-up Rate and the
Blow-up Set of Breaking Waves for a Shallow Water Equation}, Math. Z.
\textbf{233} (2000), 75--91.

\bibitem {CE2}A. Constantin and J. Escher, \textit{Particle Trajectories in
Solitary Water Waves}, Bull. Amer. Math. Soc. (N.S.) \textbf{44} (2007), 423--431.

\bibitem {CI}A. Constantin and R. Ivanov,\textit{ On an Integrable
Two-component Camassa--Holm Shallow Water System}, Phys. Lett. A \textbf{372}
(2008), 7129--7132.

\bibitem {CL}A. Constantin and D. Lannes, \textit{The Hydrodynamical Relevance
of the Camassa-Holm and Degasperis-Procesi Equations}, Arch. Ration. Mech.
Anal. \textbf{192} (2009), 165--186.

\bibitem {DXY}Y.B. Deng, J.L. Xiang and T. Yang, \textit{Blowup Phenomena of
Solutions to Euler-Poisson Equations}, J. Math. Anal. Appl. \textbf{286}
(2003), 295--306.

\bibitem {ELY2}J. Escher, O. Lechtenfeld and Z. Yin, \textit{Well-posedness
and Blow-up Phenomena for the 2-component Camassa--Holm equation}, Discrete
Contin. Dyn. Syst. Ser. A \textbf{19} (2007), 493--513.

\bibitem {GY}C.X. Guan and Z.Y. Yin, \textit{Global Existence and Blow-up
Phenomena for an Integrable Two-component Camassa--Holm Shallow Water System},
J. Differential Equations \textbf{248} (2010), 2003--2014.

\bibitem {GW}P. Goldreich and S. Weber, \textit{Homologously Collapsing
Stellar Cores}, Astrophys. J. \textbf{238 }(1980), 991--997.

\bibitem {G}Z.G. Guo, Blow-up and Global Solutions to a New Integrable Model
with Two Components, J. Math. Anal. Appl. \textbf{372} (2010), 316--327.

\bibitem {GZ}Z.G. Guo and Y. Zhou, On Solutions to a Two-component Generalized
Camassa--Holm System, Stud. Appl. Math. \textbf{124} (2010), 307--322.

\bibitem {I}R.I. Ivanov, \textit{Extended Camassa--Holm Hierarchy and
Conserved Quantities}, Z. Naturforsch. A \textbf{61} (2006), 133--138.

\bibitem {J}R. S. Johnson, \textit{Camassa-Holm, Korteweg-de Vries and Related
Models for Water Waves}, J. Fluid Mech. \textbf{455} (2002), 63--82.

\bibitem {LS}W.D. Lakin and D. A. Sanchez, Topics in Ordinary Differential
Equations, Dover Pub. Inc., New York, 1982.

\bibitem {Li}T.H. Li, \textit{Some Special Solutions of the Multidimensional
Euler Equations in }$R^{N}$, Comm. Pure Appl. Anal\textit{.} \textbf{4}
(2005), 757--762.

\bibitem {M1}T. Makino, \textit{Blowing up Solutions of the Euler-Poission
Equation for the Evolution of the Gaseous Stars}, Transport Theory and
Statistical Physics \textbf{21} (1992), 615-624.

\bibitem {SLL}M.R. Spiegel, S. Lipschutz and J. Liu, Mathematical Handbook of
Formulas and Tables, 3nd ed. New York: McGraw-Hill, 2008.

\bibitem {T}J. F. Toland, \textit{Stokes Waves}, Topol. Methods Nonlinear
Anal. \textbf{7} (1996), 1--48.

\bibitem {W}G. B. Whitham, Linear and Nonlinear Waves, Pure and Applied
Mathematics. Wiley-Interscience, New York-London-Sydney, 1974.

\bibitem {Yuen1}M. W. Yuen, \textit{Blowup Solutions for a Class of Fluid
Dynamical Equations in }$R^{N}$, J. Math. Anal. Appl. \textbf{329} (2007), 1064--1079.

\bibitem {Y1}M.W. Yuen, \textit{Analytical Blowup Solutions to the
2-dimensional Isothermal Euler-Poisson Equations of Gaseous Stars}, J. Math.
Anal. Appl. \textbf{341 (}2008\textbf{), }445--456.

\bibitem {Yuen 2}M.W. Yuen, \textit{Analytical Solutions to the Navier-Stokes
Equations}, J. Math. Phys. \textbf{49} (2008), 113102, 10pp.
\end{thebibliography}
\end{document}